\documentclass[twocolumn]{aastex631}

\definecolor{pink}{rgb}{1,0.1,.6}

\usepackage{float}
\usepackage{multirow}

\newcommand*\subtxt[1]{_{\textnormal{#1}}}

\begin{document}

\title{The TESS-Keck Survey. XXII. A sub-Neptune Orbiting TOI-1437}
\shorttitle{A sub-Neptune around TOI-1437}

\author[0000-0001-9771-7953]{Daria Pidhorodetska}
\altaffiliation{NASA FINESST Fellow}
\affiliation{Department of Earth and Planetary Sciences, University of California, Riverside, CA 92521, USA}

\author[0000-0002-0388-8004]{Emily A. Gilbert}
\affiliation{Jet Propulsion Laboratory, California Institute of Technology, 4800 Oak Grove Drive, Pasadena, CA 91109, USA}

\author[0000-0002-7084-0529]{Stephen R. Kane}
\affiliation{Department of Earth and Planetary Sciences, University of California, Riverside, CA 92521, USA}

\author[0000-0001-7139-2724]{Thomas Barclay}
\affiliation{NASA Goddard Space Flight Center, 8800 Greenbelt Road, Greenbelt, MD 20771, USA}

\author[0000-0001-7047-8681]{Alex S. Polanski}
\affil{Department of Physics and Astronomy, University of Kansas, Lawrence, KS 66045, USA}

\author[0000-0002-0139-4756]{Michelle L. Hill}
\altaffiliation{NASA FINESST Fellow}
\affiliation{Department of Earth and Planetary Sciences, University of California, Riverside, CA 92521, USA}

\author[0000-0002-3481-9052]{Keivan G.\ Stassun}
\affiliation{Department of Physics and Astronomy, Vanderbilt University, Nashville, TN 37235, USA}

\author[0000-0002-8965-3969]{Steven Giacalone}
\altaffiliation{NSF Astronomy and Astrophysics Postdoctoral Fellow}
\affiliation{Department of Astronomy, California Institute of Technology, Pasadena, CA 91125, USA}

\author[0000-0002-5741-3047]{David R. Ciardi}
\affiliation{NASA Exoplanet Science Institute - Caltech/IPAC, Pasadena, CA 91125 USA}

\author[0000-0001-6037-2971]{Andrew W. Boyle}
\affiliation{Department of Astronomy, California Institute of Technology, Pasadena, CA 91125 USA} 

\author[0000-0002-2532-2853]{Steve~B.~Howell}
\affil{NASA Ames Research Center, Moffett Field, CA 94035, USA}

\author[0000-0003-3742-1987]{Jorge Lillo-Box}
\affiliation{Centro de Astrobiolog\'ia (CAB), CSIC-INTA, ESAC campus, Camino Bajo del Castillo s/n, 28692, Villanueva de la Ca\~nada (Madrid), Spain}

\author[0000-0003-2562-9043]{Mason G. MacDougall}
\affiliation{Department of Physics \& Astronomy, University of California Los Angeles, Los Angeles, CA 90095, USA}

\author[0000-0002-3551-279X]{Tara Fetherolf}
\affiliation{Department of Physics, California State University, San Marcos, CA 92096, USA}
\affiliation{Department of Earth and Planetary Sciences, University of California, Riverside, CA 92521, USA}


\author[0000-0002-7030-9519]{Natalie M. Batalha}
\affiliation{Department of Astronomy and Astrophysics, University of California, Santa Cruz, CA 95060, USA}

\author{Ian J. M. Crossfield}
\affil{Department of Physics and Astronomy, University of Kansas, Lawrence, KS 66045, USA}

\author[0000-0001-8189-0233]{Courtney Dressing}
\affil{Department of Astronomy, University of California Berkeley, Berkeley, CA 94720, USA}

\author[0000-0003-3504-5316]{Benjamin Fulton}
\affiliation{NASA Exoplanet Science Institute/Caltech-IPAC, MC 314-6, 1200 E California Blvd, Pasadena, CA 91125, USA}

\author[0000-0001-8638-0320]{Andrew W. Howard}
\affiliation{California Institute of Technology, Pasadena, CA 91125, USA}

\author[0000-0001-8832-4488]{Daniel Huber}
\affiliation{Institute for Astronomy, University of Hawai`i, 2680 Woodlawn Drive, Honolulu, HI 96822, USA}
\affiliation{Sydney Institute for Astronomy (SIfA), School of Physics, University of Sydney, NSW 2006, Australia}

\author[0000-0002-0531-1073]{Howard Isaacson}
\affiliation{Department of Astronomy, University of California, Berkeley, Berkeley, CA 94720, USA}
\affiliation{Centre for Astrophysics, University of Southern Queensland, Toowoomba, QLD, Australia}

\author[0000-0003-0967-2893]{Erik A Petigura}
\affiliation{Department of Physics \& Astronomy, University of California Los Angeles, Los Angeles, CA 90095, USA}

\author[0000-0003-0149-9678]{Paul Robertson}
\affiliation{Department of Physics and Astronomy, University of California, Irvine, CA 92697, USA}

\author[0000-0002-3725-3058]{Lauren M. Weiss}
\affiliation{Department of Physics and Astronomy, University of Notre Dame, Notre Dame, IN 46556, USA}


\author[0000-0002-9751-2664]{Isabel Angelo}
\affiliation{Department of Physics \& Astronomy, University of California Los Angeles, Los Angeles, CA 90095, USA}

\author[0000-0001-7708-2364]{Corey Beard}
\affiliation{Department of Physics and Astronomy, University of California, Irvine, CA 92697, USA}

\author[0000-0003-0012-9093]{Aida Behmard}
\altaffiliation{Kalbfleisch Fellow}
\affiliation{Department of Astrophysics, American Museum of Natural History, 200 Central Park West, Manhattan, NY 10024, USA}

\author[0000-0002-3199-2888]{Sarah Blunt}
\affiliation{Center for Interdisciplinary Exploration and Research in Astrophysics (CIERA), Northwestern University, Evanston, IL 60208, USA}

\author[0000-0002-4480-310X]{Casey L. Brinkman}
\affiliation{Institute for Astronomy, University of Hawai`i, 2680 Woodlawn Drive, Honolulu, HI 96822, USA}

\author[0000-0003-1125-2564]{Ashley Chontos}
\affiliation{Department of Astrophysical Sciences, Princeton University, Princeton, NJ, 08544, USA}

\author[0000-0002-8958-0683]{Fei Dai}
\affiliation{Institute for Astronomy, University of Hawai`i, 2680 Woodlawn Drive, Honolulu, HI 96822, USA}
\affiliation{Division of Geological and Planetary Sciences,
1200 E California Blvd, Pasadena, CA, 91125, USA}
\affiliation{Department of Astronomy, California Institute of Technology, Pasadena, CA 91125, USA}

\author[0000-0002-4297-5506]{Paul A.\ Dalba}
\affiliation{Department of Astronomy and Astrophysics, University of California, Santa Cruz, CA 95060, USA}

\author[0000-0002-5034-9476]{Rae Holcomb}
\affiliation{Department of Physics and Astronomy, University of California, Irvine, CA 92697, USA}

\author[0000-0001-8342-7736]{Jack Lubin}
\affiliation{Department of Physics and Astronomy, University of California, Irvine, CA 92697, USA}
\affiliation{Department of Physics \& Astronomy, University of California Los Angeles, Los Angeles, CA 90095, USA}

\author[0000-0002-7216-2135]{Andrew W. Mayo}
\affil{Department of Astronomy, University of California Berkeley, Berkeley, CA 94720, USA}

\author[0000-0001-8898-8284]{Joseph M. Akana Murphy}
\altaffiliation{NSF Graduate Research Fellow}
\affiliation{Department of Astronomy and Astrophysics, University of California, Santa Cruz, CA 95060, USA}

\author[0000-0002-7670-670X]{Malena Rice}
\affiliation{Department of Astronomy, Yale University, New Haven, CT 06511, USA}

\author[0000-0003-3856-3143]{Ryan Rubenzahl}
\affiliation{Department of Astronomy, California Institute of Technology, Pasadena, CA 91125 USA} 

\author[0000-0003-3623-7280]{Nicholas Scarsdale}
\affiliation{Department of Astronomy and Astrophysics, University of California, Santa Cruz, CA 95060, USA}

\author[0000-0002-1845-2617]{Emma V. Turtelboom}
\affil{Department of Astronomy, University of California Berkeley, Berkeley, CA 94720, USA}

\author[0000-0003-0298-4667]{Dakotah Tyler}
\affiliation{Department of Physics \& Astronomy, University of California Los Angeles, Los Angeles, CA 90095, USA}

\author[0000-0002-4290-6826]{Judah Van Zandt}
\affiliation{Department of Physics \& Astronomy, University of California Los Angeles, Los Angeles, CA 90095, USA}

\author[0000-0002-2949-2163]{Edward W. Schwieterman}
\affiliation{Department of Earth and Planetary Sciences, University of California, Riverside, CA 92521, USA}

\begin{abstract}

Exoplanet discoveries have revealed a dramatic diversity of planet sizes across a vast array of orbital architectures. Sub-Neptunes are of particular interest; due to their absence in our own solar system, we rely on demographics of exoplanets to better understand their bulk composition and formation scenarios. Here, we present the discovery and characterization of TOI-1437 b, a sub-Neptune with a 18.84~day orbit around a near-Solar analog (M$_\star$ = 1.10 $\pm$ 0.10 M$_\sun$, R$_\star$ = 1.17 $\pm$ 0.12 R$_\sun$). The planet was detected using photometric data from the Transiting Exoplanet Survey Satellite (TESS) mission and radial velocity follow-up observations were carried out as a part of the TESS-Keck Survey (TKS) using both the HIRES instrument at Keck Observatory and the Levy Spectrograph on the Automated Planet Finder (APF) telescope. A combined analysis of these data reveal a planet radius of R$_p$ = 2.24 $\pm$ 0.23 R$_\earth$ and a mass measurement of M$_p$ = 9.6 $\pm$ 3.9 M$_\earth$). TOI-1437 b is one of few ($\sim$50) known transiting sub-Neptunes orbiting a solar-mass star that has a radial velocity mass measurement. As the formation pathway of these worlds remains an unanswered question, the precise mass characterization of TOI-1437 b may provide further insight into this class of planet.
\end{abstract}



\section{Introduction}
\label{sec:intro}

In the ever-expanding field of exoplanetary research, each discovery marks a stride towards unraveling the mysteries of distant worlds. The CoRoT \citep{leger2009}, $Kepler$ \citep{Borucki2010}, and K2 missions \citep{howell2014} paved the way for space-based detection of planets smaller than Neptune ($\lesssim4 \mathrm{R}_\earth$) through the use of the transit method. Although $Kepler$ identified a plethora of sub-Neptunes within our galaxy \citep{Dressing2013,Fulton2018}, relatively few of these planets were suitable radial velocity (RV) follow-up targets as a result of their large distances. 

The formation \citep[e.g.][]{Rogers2011,Schlichting2014,Raymond2018}, atmospheric composition \citep[e.g.][]{Morley2017,Kempton2018}, and interior structure \citep[e.g.][]{GuzmanMesa2022,Misener2023} of sub-Neptunes are not well understood. These planets, ranging from 1-4 R$_\earth$, are expected to vary in composition from rocky to gas-rich \citep{Fortney2013,Moses2013,lopez2014}, while planets with radii larger than $~$1.6 R$_\earth$ retain hydrogen envelopes \citep{Rogers2015,Fulton2017}. The mass-radius relation can provide the initial assessment of how likely it is that the planet hosts an atmosphere, but these parameters are often unknown. Although around a third of Sun-like stars host sub-Neptunes with orbital periods of less than 100 days \citep{Fressin2013,Petigura2013,Burke2015,Hsu2019}, relatively few ($\sim$50) of these planets have known mass measurements. This strengthens the need for RV observations of these worlds to further understand their nature.

The Transiting Exoplanet Survey Satellite \citep[\text{TESS};][]{Ricker2015} has added to the exoplanet inventory by carrying out an all-sky survey attempting to detect small, transiting planets around bright, nearby stars. Revealing over 6,000 planet candidates thus far, many of these TESS Objects of Interest (TOIs) are amenable to follow-up with RV measurements to determine planetary masses \citep{Guerrero2021,Kane2021a}. The TESS-Keck Survey \citep[\text{TKS};][]{Chontos2022}, a large monitoring program of TESS planet candidates with the Keck-HIRES and APF-Levy Spectrographs, is working to provide the precise planet mass measurements required by future efforts targeting atmospheric characterization that often include sub-Neptunes (e.g. \citealt{Scarsdale2021,Lubin2022,Lange2024}).  Measurements of planet radii in combination with masses have proven to be the most advantageous method for understanding the compositions of exoplanets by allowing us to discern the bulk densities of planets \citep{howe2014,lopez2014,wolfgang2016,unterborn2023}. These combined measurements are particularly important for sub-Neptune exoplanets that have no analog representation within the solar system \citep{Kane2021b}.

Here, we announce the discovery and precise mass measurement of a planet orbiting TOI-1437 (otherwise known as HD 154840; BD+57 1730; TIC 198356533), a near-Solar analog (M$_\star$ = 1.10 $\pm$ 0.10 M$_\sun$, R$_\star$ = 1.17 $\pm$ 0.12 R$_\sun$) observed by TESS. Section~\ref{sec:obs} describes the observations and collected data of the system including photometry, spectroscopy, and imaging. Section~\ref{sec:stellar} details the host star properties, followed by an analysis of the data modeling in Section~\ref{sec:analysis}. A discussion of the results and findings is found in Section~\ref{sec:discussion} with concluding remarks in Section~\ref{sec:conclusions}.

\section{Observations}
\label{sec:obs}

\subsection{Photometry}
\label{sec:phot}

\begin{figure}
    \includegraphics[width=0.5\textwidth]{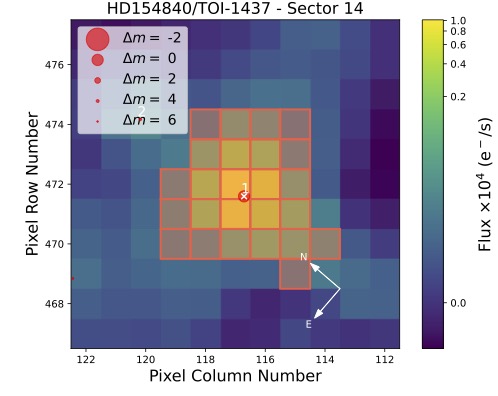}
    \caption{Target pixel file corresponding to TOI-1437 observations by TESS in Sector 14. Sources from $Gaia$ DR3 are shown as red circles, with their size corresponding to the magnitude contrast against TOI-1437 (marked with the label "1" and a white cross). The aperture used by the TESS-SPOC pipeline is shown as a shaded red region in each panel.}\label{fig:TPF}  
\end{figure}

TOI-1437 lies within the northern Continuous Viewing Zone which has been observed exhaustively by TESS over its full mission lifetime. TESS observed TOI-1437 in Year 2 of operations at 2-minute cadence during Sectors 14-26. TOI-1437 was also observed in Sectors 40, 41, and 47-60 at both 20-second and 2-minute cadence. TOI-1437 will be re-observed by TESS in upcoming Sectors 73-83. The time-series observations of TOI-1437 were processed with the TESS Science Processing Operations Center (SPOC) pipeline \citep{Jenkins2016} which resulted in the detection of a periodic transit signal for TOI-1437.01 that was alerted on 2019 November 14. No significant photometric variability is present in the light curve data. \

To check for sources of contamination within the pipeline aperture, we use the \texttt{tpfplotter} algorithm\footnote{\url{https://github.com/jlillo/tpfplotter}} to inspect the target pixel files (TPFs) for each sector using the procedure described in \citealt{Aller2020}. Figure \ref{fig:TPF} shows the TPF of Sector 14 overlaid with the $Gaia$ Data Release 3 (DR3) catalog \citep{Vallenari2023} consisting of bright sources down to 6 mag more faint than TOI-1437. The TPF shows no indication of any sources of contamination within the aperture of TESS.


\subsection{Spectroscopy}
\label{sec:spectroscopy}

\subsubsection{HIRES Observations}

TOI-1437 was observed over 88 nights with the HIRES instrument at the Keck Observatory \citep{Vogt1994} between UT 2019 November 28 and UT 2023 June 26, including one high signal-noise-ratio (SNR) iodine-free template spectrum. HIRES is an echelle spectrometer which observes over a spectral range of 3000-6500 {\AA}. HIRES uses the iodine technique, where the wavelength range for RV measurements is 5000-6300 {\AA}. As there are no prominent telluric lines in the wavelength section used in the iodine-cell technique, they are not masked out. For microtellurics (small absorption features caused by the atmosphere) or lines such as the Sodium D lines that can be in emission due to street light reflection in the atmosphere, the iodine technique implements a weighting technique that gives higher weights to the spectral segments that provide less scatter through an observation stack for each star. This effectively de-weights wavelength sections with problematic telluric features. Our HIRES RVs have a median binned uncertainty of 1.49 m/s and a median exposure time of 329 seconds. We reduced the RVs using the standard procedure described in \citet{Howard2010}. The first ten HIRES RV measurements are provided in Table \ref{tab:rvmeasurements} while a complete list is available online.

\vfill\null
\subsubsection{APF Observations}

In addition to the Keck/HIRES observations, we collected 106 spectra of TOI-1437 with the Levy Spectrograph on the Automated Planet Finder (APF) telescope \citep{Vogt2014} between UT 2019 December 13 and 2022 July 30. The APF is a 2.4-m telescope located at the Mt. Hamilton station of UCO/Lick Observatory. The telescope is coupled with the high resolution (R$\subtxt{max}$ $\sim$120,000) prism cross-dispersed Levy echelle spectrograph, which covers a wavelength range of $\sim$3700--9700 {\AA}. APF uses the same iodine technique as HIRES, taking RV measurements between 5000-6300 {\AA} and negating the need to mask out telluric lines. Our APF RVs have a median binned uncertainty of 5.22 m/s and a median exposure time of 1800 seconds. The APF/Levy Doppler software was developed based on the Keck/HIRES Doppler software and therefore follows a similar process for reducing spectra to RVs. A full description of the design and individual components of the APF is available in \citealt{Vogt2014}. The first ten APF RV measurements are provided in Table \ref{tab:rvmeasurements}. 


\begin{table*}[t]
\centering 
\caption{Radial Velocity Measurements} 
\begin{tabular}{c c c c c }
\hline
\hline
Time (BJD) & RV (m s$^{-1}$) & RV Unc. (m s$^{-1}$) & S$_{HK}$ & Instrument  \\
\hline
2458831.077696 & -2.23 & 5.72 & 0.1450 & Levy \\ 
2458831.091608 & 7.68 & 6.27 & 0.1375 & Levy \\ 
2458852.10649 & 21.07 & 6.69 & 0.1171 & Levy \\
2458876.070009 & -12.17 & 7.14 & 0.1286 & Levy \\ 
2458876.084558 & -1.64 & 7.14 & 0.1337 & Levy \\ 
2458882.049758 & -4.89 & 10.49 & 0.1440 & Levy \\ 
2458883.941799 & 14.17 & 12.18 & 0.0974 & Levy \\ 
2458885.042157 & -0.77 & 10.22 & 0.3082 & Levy \\ 
2458885.87575 & 24.68 & 10.37 & 0.1557 & Levy \\ 
2458887.016514 & 1.33 & 6.57 & 0.1255 & Levy \\
2458917.092394 & 0.54 & 1.90 & 0.1204 & HIRES \\ 
2458918.07703 & 3.10 & 1.73 & 0.126 & HIRES \\ 
2458999.90562 & -0.34 & 1.62 & 0.1301 & HIRES \\ 
2459002.948851 & 1.34 & 1.57 & 0.1289 & HIRES \\
2459003.911424 & 0.81 & 1.55 & 0.1304 & HIRES \\
2459006.907673 & 1.43 & 1.36 & 0.1308 & HIRES \\
2459007.899261 & 1.39 & 1.36 & 0.1301 & HIRES \\ 
2459010.936502 & -1.20 & 1.43 & 0.1305 & HIRES \\
2459011.850521 & 0.09 & 1.37 & 0.1316 & HIRES \\ 
2459012.801402 & 4.34 & 1.49 & 0.1335 & HIRES\\
\hline
\end{tabular}\\[5pt]Note: Only the first 10 APF/Levy and Keck/HIRES RVs are displayed in this table. A complete list has been made available online. S$\subtxt{HK}$ values were measured using procedures from \cite{Isaacson2010} with standard uncertainties of 0.002 for APF/Levy measurements and 0.001 for Keck/HIRES measurements.\label{tab:rvmeasurements}\end{table*} 


\subsection{Imaging}
\label{sec:imaging}

Close stellar companions (bound or line of sight) can confound exoplanet discoveries in a number of ways.  The detected transit signal might be a false positive due to a background eclipsing binary, and even real planet discoveries will yield incorrect stellar and exoplanet parameters if a close companion exists and are unaccounted for \citep{Ciardi2015,FurlanHowell2020}. Additionally, the presence of a close companion star leads to the non-detection of small planets residing with the same exoplanetary system \citep{Lester2021}. Given that nearly one-half of solar-like stars are in binary or multiple star systems \citep{Matson2018}, high-resolution imaging provides crucial information toward our understanding of exoplanetary formation, dynamics and evolution \citep{Howell2021}. \

As part of our standard process for validating transiting exoplanets to assess the the possible contamination of bound or unbound companions on the derived planetary radii \citep{Ciardi2015}, we observed TOI-1437 with high-resolution near-infrared adaptive optics (AO) imaging at Palomar and Lick Observatories and in the optical with lucky imaging at Calar Alto and speckle imaging at Gemini-North. The infrared observations provide the deepest sensitivities to faint companions, while the optical speckle observations provide the highest resolution imaging, making the two techniques complementary. 


\subsubsection{Palomar AO}

\begin{figure}
    \centering
    \includegraphics[width=0.52\textwidth]{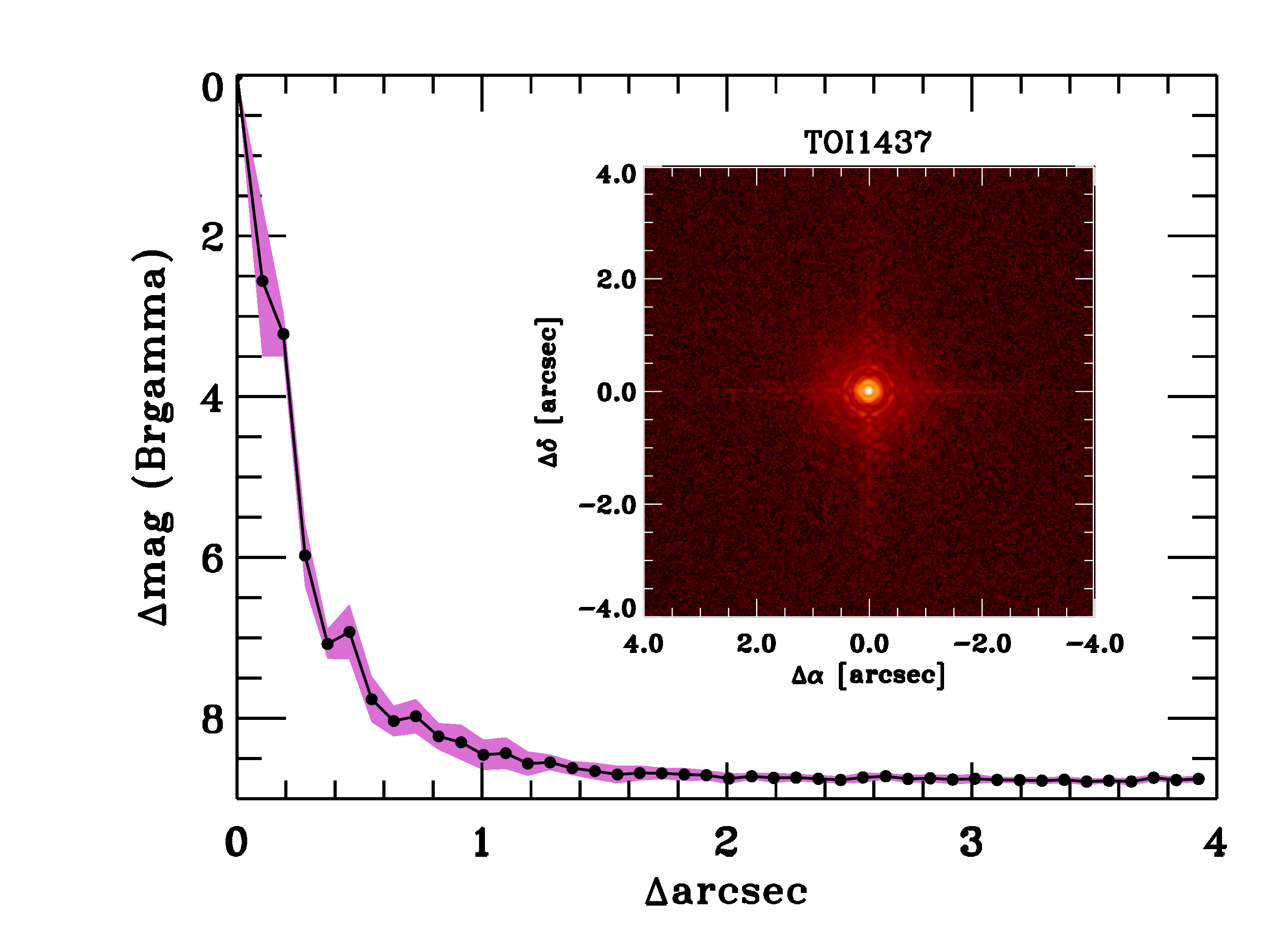}
    \caption{Companion sensitivity for the near-infrared adaptive optics imaging.  The black points represent the 5$\sigma$ limits and are separated in steps of 1 FWHM; the purple represents the azimuthal dispersion (1$\sigma$) of the contrast determinations. The inset image is of the primary target showing no additional close-in companions.}\label{fig:ao_contrast}  
\end{figure}

The Palomar Observatory observations were made with the PHARO instrument \citep{hayward2001} behind the natural guide star AO system P3K \citep{dekany2013} on UT 2021~June~22  in a standard 5-point quincunx dither pattern with steps of 5\arcsec\ in the narrow-band $Br-\gamma$ filter $(\lambda_o = 2.1686; \Delta\lambda = 0.0326~\mu$m).  Each dither position was observed three times, offset in position from each other by 0.5\arcsec\ for a total of 15 frames; with an integration time of 1.4 seconds per frame, the total on-source time was 21 seconds. PHARO has a pixel scale of $0.025\arcsec$ per pixel for a total field of view of $\sim25\arcsec$. The science frames were flat-fielded and sky-subtracted. The reduced science frames were combined into a single image with a final resolution of 0.091\arcsec FWHM (Figure~\ref{fig:ao_contrast}). \ 

To within the limits of the AO observations, no stellar companions were detected. The sensitivities of the final combined AO image were determined by injecting simulated sources azimuthally around the primary target every $20^\circ $ at separations of integer multiples of the central source's FWHM \citep{furlan2017a, lund2020}. The brightness of each injected source was scaled until standard aperture photometry detected it with $5\sigma $ significance. The resulting brightness of the injected sources relative to TOI-1437 set the contrast limits at that injection location. The final $5\sigma $ limit at each separation was determined from the average of all of the determined limits at that separation, and the uncertainty on the limit was set by the rms dispersion of the azimuthal slices at a given radial distance.

\subsubsection{Shane AO}
We observed TOI-1437 on UT 2021 March 05 using the ShARCS camera on the Shane 3-meter telescope at Lick Observatory \citep{2012SPIE.8447E..3GK, 2014SPIE.9148E..05G, 2014SPIE.9148E..3AM}. Observations were taken with the Shane adaptive optics system in natural guide star mode in order to search for nearby, unresolved stellar companions. We collected sequences of observations using a $K_s$ filter ($\lambda_0 = 2.150$ $\mu$m, $\Delta \lambda = 0.320$ $\mu$m) and a $J$ filter ($\lambda_0 = 1.238$ $\mu$m, $\Delta \lambda = 0.271$ $\mu$m). We reduced the data using the publicly available \texttt{SImMER} pipeline \citep{2020AJ....160..287S, 2022PASP..134l4501S}.\footnote{https://github.com/arjunsavel/SImMER} Our observations achieve contrasts of 4.5 ($K_s$) and 4.0 ($J$) at 1$\arcsec$. We find no nearby stellar companions within our detection limits.


\subsubsection{Gemini Speckle}
\begin{figure}
    \centering
    \includegraphics[width=0.51\textwidth]{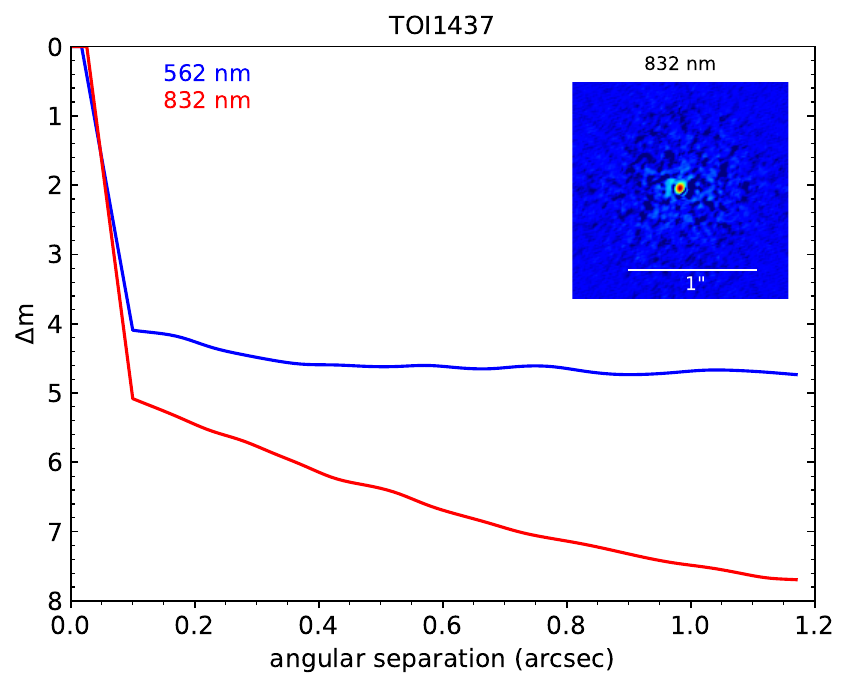}
    \caption{Companion sensitivity for speckle imaging. Contrast curves (5$\sigma$) for 562 nm are shown in blue and 832 nm in red. The reconstructed 832 nm speckle image (top right) shows that no close companions to TOI-1437 were detected within the obtained contrast and angular limits.}\label{fig:geminispeckle}  
\end{figure}

TOI-1437 was observed on UT 2020 February 18 using the ‘Alopeke speckle instrument on the Gemini North 8-m telescope \citep{Scott2021}. ‘Alopeke provides simultaneous speckle imaging in two bands (562 nm and 832 nm) with output data products including a reconstructed image with robust contrast limits on companion detections \citep{HowellFurlan2022}. Three sets of 1000 x 0.06 second images were obtained and processed in our standard reduction pipeline (see \citealt{Howell2011}). We find that TOI-1437 is a single star with no close companions brighter then 5--7.5 magnitudes below that of the target star and within the angular and contrast limits achieved (Figure \ref{fig:geminispeckle}). The angular limits, the Gemini 8-m telescope diffraction limit (20 mas) out to 1.2”, correspond to spatial limits of 2 to 124 AU at the distance of TOI-1437 (d=103 pc). The 562 and 832 nm bands allow us to rule out close companions brighter than an M2V or an M4V, respectively.  \ 
 
\subsubsection{Calar Alto Lucky}
We obtained a high-spatial resolution image of TOI-1437 using the AstraLux instrument \citep{Hormuth2008}, located at the 2.2\,m telescope of the Calar Alto Observatory (Almer\'{\i}a, Spain). The observations were executed on the night of UT 2020 February 25 under moderately poor weather conditions (average seeing of 1.5 arcsec) and at an airmass of 1.08. We obtained 54,457 frames in the SDSSz bandpass with an individual exposure time of 10\,ms and a field-of-view windowed to $6\times6$ arcsec. The datacube was then processed by the automatic observatory pipeline \citep{Hormuth2008}, which besides doing the basic reduction of the individual frames, selects the 10\% with the highest Strehl ratio \citep{Strehl1902} and combines them into a  final high-spatial resolution image. This final image does not show any additional companions within the sensitivity limits. Such limits are computed using our own developed \texttt{astrasens} package\footnote{\url{https://github.com/jlillo/astrasens}} with the procedure described in \citealt{lillo-box12,LilloBox2014}.

\begin{figure}
    \centering
    \includegraphics[width=0.53\textwidth]{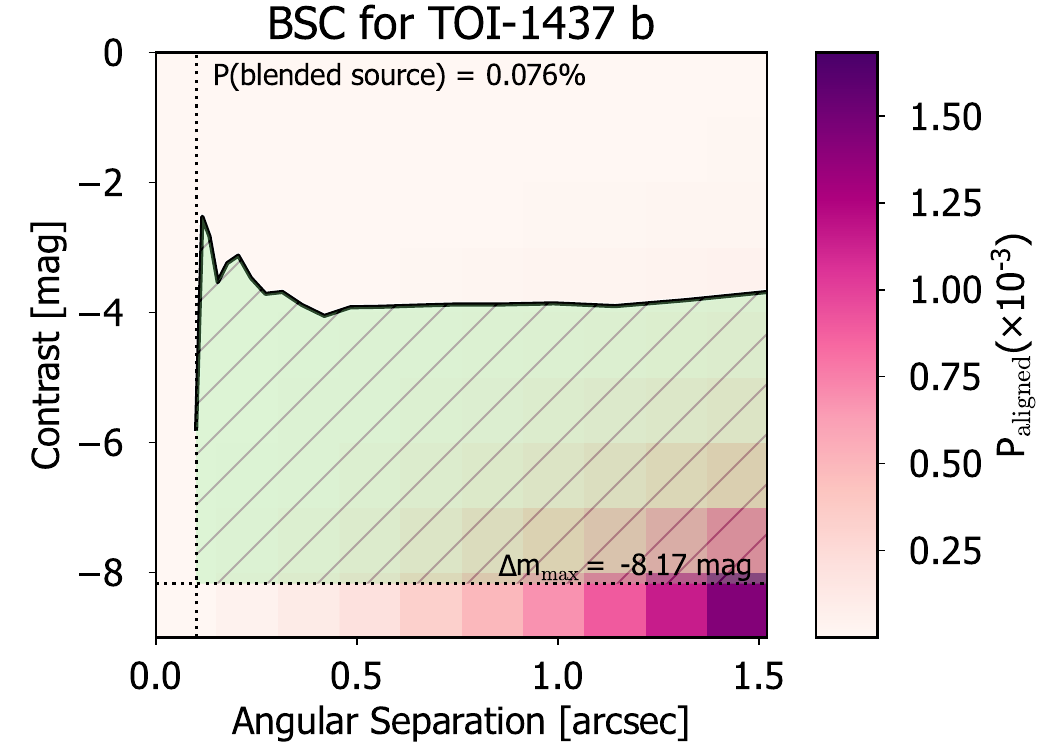}
    \caption{Results of the blended source confidence (BSC) analysis of Calar Alto Lucky imaging data indicate that the probability for TOI-1437 to have a blended, undetected, eclipsing binary capable of mimicking our transit is very low (0.076\%).}\label{fig:astralux}  
\end{figure}

Subsequently, we used this contrast curve to establish the probability that the observed signal came from a chance-aligned eclipsing binary with capabilities of producing a transit depth mimicking the one observed and that could have been missed by our high-spatial resolution image. We call this probability the blended source confidence (BSC) and the steps for estimating it are fully described in \cite{LilloBox2014}. We use a Python implementation of this approach (\texttt{bsc}, by J. Lillo-Box) which uses the TRILEGAL\footnote{\url{http://stev.oapd.inaf.it/cgi-bin/trilegal}} galactic model \citep[v1.6;][]{girardi12} to retrieve a simulated source population of the region around the corresponding target\footnote{This is done in python by using the \texttt{astrobase} implementation by \cite{astrobase}.}. We used the same parameters as in previous works (e.g., \citealt{bluhm21,soto21}), namely the default parameters for the bulge, halo, thin/thick disks, and the lognormal initial mass function from \cite{chabrier01}. The results from the BSC analysis are shown in Figure~\ref{fig:astralux} and show a very low probability of 0.076\% for our target to have a blended, undetected, eclipsing binary capable of mimicking our transit. Consequently, we assume from here on that the transit is associated with TOI-1437 and that it is not caused by any other source.

\section{Stellar Characterization}
\label{sec:stellar}

\subsection{Gaia Assessment}
In addition to the high resolution imaging, we have utilized $Gaia$ to identify any wide stellar companions that may be bound members of the system.  Typically, these stars are already in the TESS Input Catalog and their flux dilution to the transit has already been accounted for in the transit fits and associated derived parameters.  Based upon similar parallaxes and proper motions \citep[e.g.,][]{mugrauer2020,mugrauer2021,mugrauer2022}, there are no additional widely separated companions identified by $Gaia$.

Additionally, the $Gaia$ DR3 astrometry provides information on the possibility of inner companions that may have gone undetected by either $Gaia$ or the high resolution imaging. The $Gaia$ Renormalised Unit Weight Error (RUWE) is a metric, similar to a reduced chi-square, where values that are $\lesssim 1.4$  indicate that the $Gaia$ astrometric solution is consistent with the star being single whereas RUWE values $\gtrsim 1.4$ may indicate an astrometric excess noise, possibily caused the presence of an unseen companion \citep[e.g., ][]{ziegler2020}.  TOI-1437 has a $Gaia$ DR3 RUWE value of 0.996 indicating that the astrometric fits are consistent with the single star model. 


\subsection{Stellar Properties}
\begin{figure}
    \centering    
    \includegraphics[width=0.5\textwidth,trim=80 70 50 50,clip]{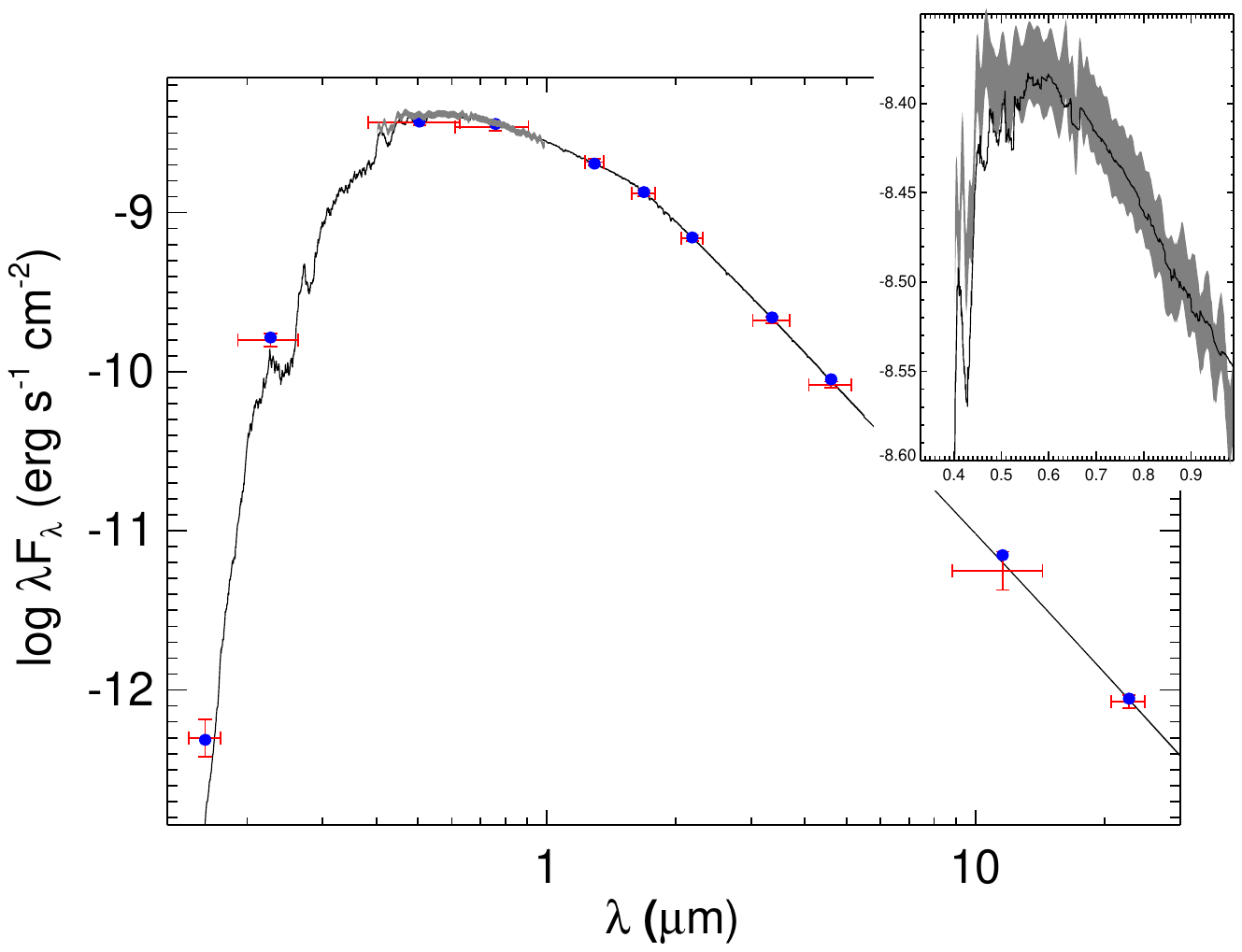}
\caption{Spectral energy distribution of TOI-1437. Red symbols represent the observed photometric measurements, where the horizontal bars represent the effective width of the passband. Blue symbols are the model fluxes from the best-fit PHOENIX atmosphere model (black). The inset shows the absolute flux-calibrated {\it Gaia\/} spectrum as a gray swathe overlaid on the model. \label{fig:sed}}
\end{figure}

As an independent determination of the basic stellar parameters, we performed an analysis of the broadband spectral energy distribution (SED) of the star together with the {\it Gaia\/} DR3 parallax \citep[with no systematic offset applied; see, e.g.,][]{StassunTorres:2021}, in order to determine an empirical measurement of the stellar radius, following the procedures described in \citet{Stassun:2016,Stassun:2017,Stassun:2018}. We pulled the $JHK_S$ magnitudes from {\it 2MASS}, the W1--W4 magnitudes from {\it WISE}, the $G_{\rm BP} G_{\rm RP}$ magnitudes from $Gaia$ DR3, and the FUV and NUV magnitudes from {\it GALEX}. We adopted a minimum uncertainty of 0.03~mag to account for known systematics in the absolute flux calibration of the ground-based photometric systems \citep[see, e.g.,][]{Stassun:2016}. We also utilized the absolute flux-calibrated low-resolution {\it Gaia\/} spectra. Together, the available photometry spans the full stellar SED over the wavelength range 0.2--20~$\mu$m (see Figure~\ref{fig:sed}).  

We performed a fit using PHOENIX stellar atmosphere models \citep{Husser:2013}, with the principal parameters being the effective temperature ($T_{\rm eff}$) and metallicity ([Fe/H]), for which we adopted the spectroscopically determined values, as well as the extinction $A_V$, which we limited to the maximum line-of-sight value from the Galactic dust maps of \citet{Schlegel1998}. The resulting model (Figure~\ref{fig:sed}) has a best-fit $A_V = 0.04 \pm 0.02$ with a reduced $\chi^2$ of 0.8. 

Integrating the (unreddened) model SED gives the bolometric flux at Earth, $F_{\rm bol}$, which with the {\it Gaia\/} parallax directly yields the bolometric luminosity, $L_{\rm bol}$. Taking the $L_{\rm bol}$ and $T_{\rm eff}$ together provides us with the stellar radius, $R_\star$. This radius, along with the determined log \textit{g}, allows us to derive a stellar mass that is consistent with the \cite{Torres:2010} relations. A full list of the characterized stellar parameters is presented in Table \ref{tab:properties}.



\begin{table}
\centering 
\caption{TOI-1437 Stellar Properties} 
\begin{tabular}{c c l}
\hline
\hline
Parameter & Value & Notes \\
\hline
RA (\textdegree) & 256.1387540990 & A \\
Dec (\textdegree) & 56.8425461418 & A \\ 
$\mu_{\alpha}$ (mas/yr) & 10.799 $\pm$ 0.013 & A \\
$\mu{\delta}$ (mas/yr) & -20.284 $\pm$ 0.015 & A \\
$\varpi$ (mas) & 9.667 $\pm$ 0.010 & A \\ 
{[Fe/H]} & -0.4924 $\pm$ -0.4983 & A \\ 
V (mag) & 9.173 $\pm$ 0.003 & B \\
B (mag) & 9.749 $\pm$ 0.042 & B \\
E$_{(B-V)}$ &  0.0059 $\pm$ 0.0027 & B \\ 
T$\subtxt{eff}$ (K) & 6008 $\pm$ 113 & C \\
Luminosity (L$_\sun$) & 1.926 $\pm$ 0.023 & C\\
F$_{\rm bol}$  (erg~s$^{-1}$~cm$^{-2})$ & 5.774 $\pm 0.068 \times 10^{-9}$  & C \\
log \textit{g} & 4.26 $\pm$ 0.04 & C \\
Mass (M$_\sun$) & 1.10 $\pm$ 0.10 & C \\ 
$\rho$$_*$ (g/cc) & 0.5023904 & C \\

 
\hline
\hline
\end{tabular}
\\[5pt]A: \textit{Gaia} DR3 \citep{Vallenari2023}; B: TIC v8.2; C: SED analysis \ 
\label{tab:properties}
\end{table}

\section{Analysis}
\label{sec:analysis}

\subsection{Photometric and RV Analysis}
\begin{figure*}
    \centering
    \includegraphics[width=1.03\textwidth]{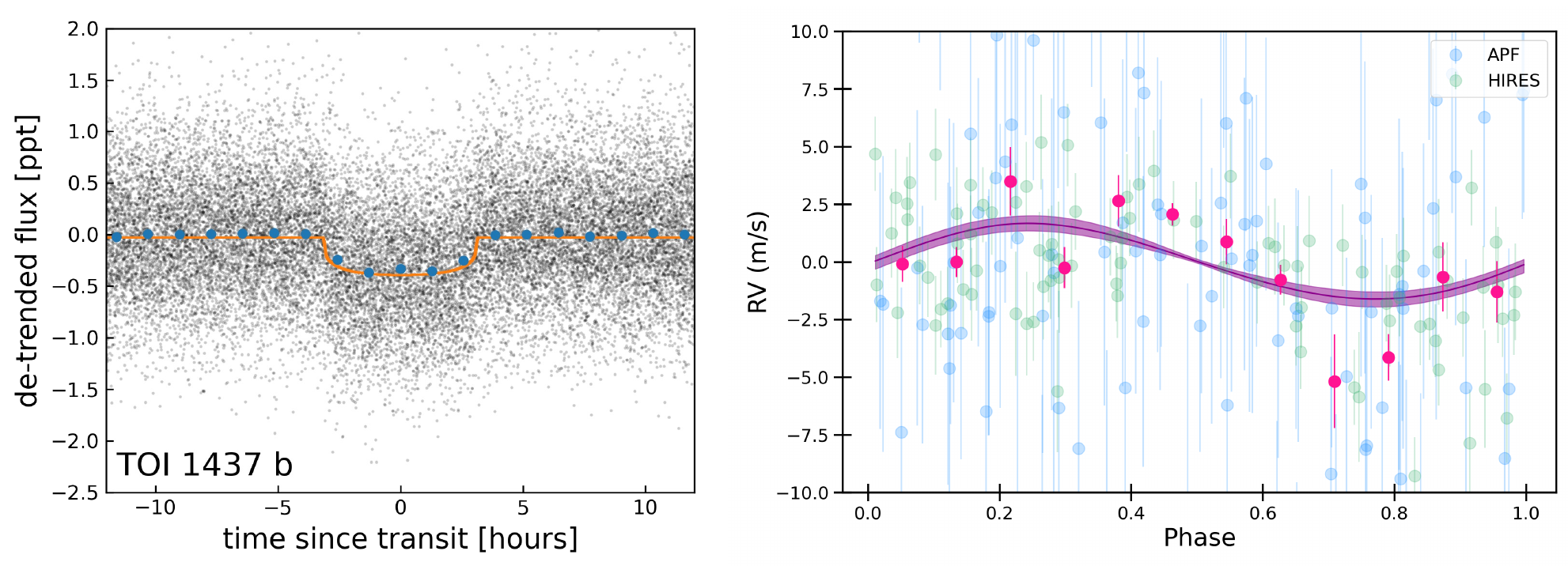}
    \caption{Left: Transit model drawn from parameter posterior distributions fit from phase-folded TESS photometry of TOI-1437. The grey dots show the TESS 2 minute cadence data. The orange line shows the best fit to each individual transit using \texttt{exoplanet}. The blue dots show the data binned to 2 minute cadence. Right: Phase-folded views of the best-fit RV model for TOI-1437 b (purple). Data from APF is shown in blue while HIRES is displayed in green, and each is binned to 12 total bins.}
    \label{fig:rv_fit}
\end{figure*}

A joint model for the TESS light curve data and the two RV datasets from HIRES and APF (Figure \ref{fig:rv_fit}) was constructed using the framework provided by the \texttt{exoplanet} package \citep{ForemanMackey2021}. All available 2-minute cadence data (Section \ref{sec:phot}) was used. For this analysis, we did not include the 20-s cadence data. We normalized each sector to the median flux, centered on zero and combined the data into a single time series where the baseline was removed. We then sigma clipped the data for outliers more than 7-$\sigma$ from the data median. All but one of the 20 clipped points were positive outliers. We also put these data on a time axis centered on zero (e.g. the center of the dataset) as this can reduce correlations between orbital period and transit epoch.
For the RV data, the two instruments were treated as independent data sets. Each set was centered on median zero (m/s) and placed on the same time coordinates as the light curve data. \ 

The model used Gaussian Processes (GPs) to account for correlated noise in both the RV data and the light curve, and the GP hyperparameters are the same for the two RV datasets. The GP used is a non-periodic simple harmonic oscillator implemented using the celerite2 software \citep{exoplanet:foremanmackey17, exoplanet:foremanmackey18}. To avoid overfitting, we also modeled the RV data without the use of a GP and determined that a non-GP model fit gives consistent results with the fit including a GP. \ 

The complete set of model parameters is listed in Table~\ref{tab:modelparameters}. Of note, we sampled eccentricity in as a unit disk which minimizes geometric biases in sampling and also included a physical prior based on \citet{exoplanet:kipping13}. \texttt{exoplanet} uses a gradient-based MCMC algorithm that is a generalization of the No U-Turn Sampling method \citep{Hoffman2011,Betancourt2016} implemented in pymc3 \citep{exoplanet:pymc3,exoplanet:theano}. We ran the sampler for 8000 tuning samples and then drew 2000 samples in each of 4 independent chains for a total of 8000 samples. The model showed strong agreement between the 4 chains, indicating excellent convergence. The complete set of modeled outputs is found in Table \ref{tab:results}.

\begin{table}[t]
\centering 
\caption{Modeled Parameters and Priors} 
\begin{tabular}{l l}
\hline
\hline
Parameter & Value/Source  \\
\hline
Stellar radius  (R$_\odot$)& N(1.1765, 0.114)\\
Stellar density ($\rho_\odot$)& LN(ln(0.99), 0.25)\\
Orbital period (days) & N(18.841,0.001)\\
Transit epoch (BJD) & N(2459906.55,0.02)\\
Planet-to-star radius ratio & LN(ln(0.0003), 2)\\
Impact parameter & U(0,1+($r/R_\star$))\\
Limb darkening u$_1$& \citet{exoplanet:kipping13}\\
Limb darkening u$_2$& \citet{exoplanet:kipping13}\\
Orbital eccentricity & UnitDisk/\citet{Kipping2013}\\
Periastron angle (rad)& UnitDisk\\
RV semiamplitude (m~s$^{-1}$)& LN(ln(2), 5)\\
TESS jitter (ppt)& InverseGamma(93,41)\\
TESS zeropoint (ppt)& N(0,1)\\
TESS GP $\rho$ (days)& InverseGamma(3.3,5.3)\\
TESS GP $\sigma$ (ppt)& InverseGamma(93,41)\\
HIRES jitter (m~s$^{-1}$)& InverseGamma(1.5,0.6)\\
HIRES zeropoint (m~s$^{-1}$)& N(0,3)\\ 
APF jitter (m~s$^{-1}$)& InverseGamma(0.8,0.04) \\
APF zeropoint (m~s$^{-1}$)& N(0,3)\\ 
RV GP $\rho$ (days)& InverseGamma(3,40)\\
RV GP $\sigma$ (m~s$^{-1}$)& InverseGamma(0.8,0.04)\\
\hline
\hline
\end{tabular}
\label{tab:modelparameters}
\end{table}

\begin{table*}[!ht]
\centering 
\caption{Modeling Results} 
\begin{tabular}{l l l l}
\hline
\hline
Parameter & Median & +1$\sigma$ & -1$\sigma$  \\
\hline
Stellar radius  (R$_\odot$)& 1.17 & 0.11 & 0.12\\
Stellar density ($\rho_\odot$)& 0.91 & 0.23 & 0.18\\
Orbital period (days) & 18.840942 & 0.000060 & 0.000068\\
Transit epoch (BJD) & 2459360.1652 & 0.0012 & 0.0012\\
Impact parameter & 0.24 & 0.24 & 0.17\\
Planet-to-star radius ratio & 0.01746 & 0.00047 & 0.00039\\
Limb darkening u$_1$& 0.19 & 0.21 & 0.14\\
Limb darkening u$_2$& 0.57 & 0.22 & 0.30\\
Orbital eccentricity & 0.17 & 0.11 & 0.09\\
Periastron angle (rad)& -1.68 & 0.57 & 0.45\\
Radial velocity semiamplitude (m~s$^{-1}$)& 2.24 & 0.68 & 0.71\\
TESS jitter (ppt)& 0.4222 & 0.0024 & 0.0024\\
TESS GP $\rho$ (days)& 1.02 & 0.22 & 0.17\\
TESS GP $\sigma$ (ppt)& 0.0747 & 0.0058 & 0.005\\
HIRES jitter (m~s$^{-1}$)& 0.41 & 0.61 & 0.21 \\
APF jitter (m~s$^{-1}$)& 4.3 & 1.2 & 1.4\\
RV GP $\rho$ (days)& 6.1 & 3.0 & 2.0\\
RV GP $\sigma$ (m~s$^{-1}$)& 4.18 & 0.43 & 0.38\\
Planetary radius (R$_\oplus$) & 2.24 & 0.23 & 0.22\\
Planetary mass (M$_\oplus$) & 9.6 & 3.9 & 3.3 \\ 
Semimajor axis to stellar radius ratio & 25.7 & 2.0 & 1.8\\
Semimajor axis (AU) & 0.140 & 0.018 & 0.017\\ 
Orbital inclination(deg)& 89.53 & 0.33 & 0.46\\
Transit duration (hr) & 4.65 & 0.45 & 0.50\\
Insolation (S$_\oplus$) & 82.0 & 14.0 & 12.0\\
\hline
\hline
\end{tabular}
\label{tab:results}
\end{table*}




\section{Discussion}
\label{sec:discussion}
\begin{figure*}
    \centering
    \includegraphics[width=0.7\textwidth]
    {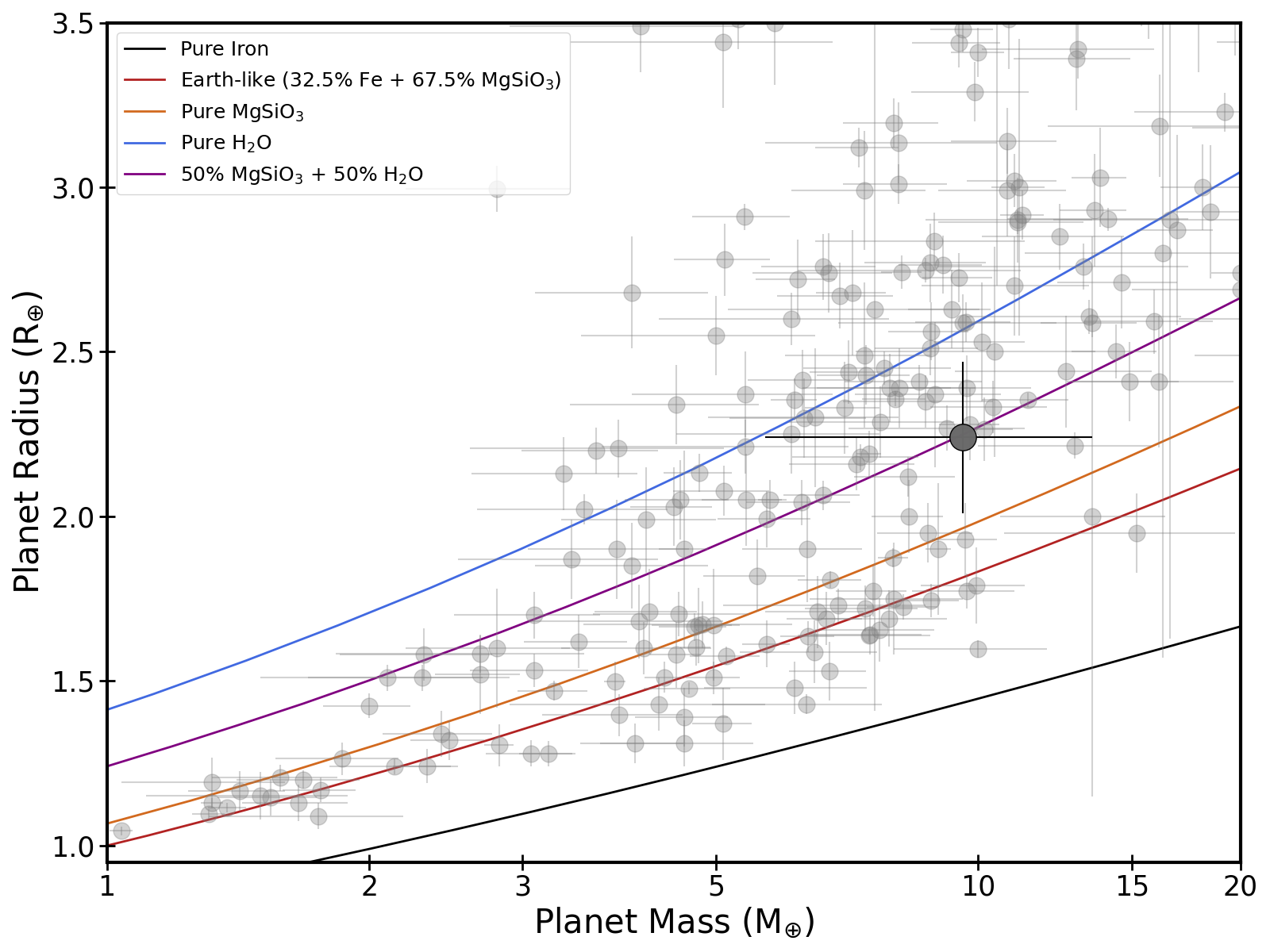}
    \caption{Representation of TOI-1437 b as it falls on a mass-radius diagram (large black circle). Individual solid lines are theoretical mass-radius curves for planets whose composition is specified in the figure legend. Light grey circles indicate the population of planets between 1.0-3.5 R$_\oplus$ that have measured masses.}\label{fig:mass-radius}
\end{figure*}

Determining the atmospheric composition of a sub-Neptune such as TOI-1437 b would provide insight into some of the many processes of these planets that are not well understood, such as their formation and interior composition. Shown in Figure~\ref{fig:mass-radius} is a mass-radius diagram, where known planets are represented as grey circles, and TOI-1437 b is indicated by the large black circle. Due to their wide range of equilibrium temperatures \citep[most falling between 400-1200 K;][]{Crossfield2017}, sub-Neptunes are thought to have a large range of atmospheric compositions \citep{kite2020}. Although larger (2.5--4.0~R$_\earth$) sub-Neptunes are expected to retain most of their primordial H/He envelopes, intermediate worlds (1.5--2.5~R$_\earth$), such as TOI-1437 b, are expected to be extremely diverse in atmospheric composition. Planets with R$_p$ $>$ 2.0 R$_\earth$ could be water-ocean worlds that contain dense steam atmospheres \citep{bean2021} or worlds hosting deep magma oceans \citep{kite2020} resulting in H$_2$- or H$_2$O-rich envelopes that are shrouded with silicate species \citep{Schlichting2022}. Based on its location within Figure~\ref{fig:mass-radius}, TOI-1437 b falls directly on the composition line of 50\% MgSiO$_3$ + 50\% H$_2$O. Although further observations would be necessary to determine this, TOI-1437 b is likely a volatile-rich planet that lacks an extended hydrogen atmosphere, but could be dominated by other gases such as CO$_2$.\

Additionally, our measurements are consistent with a single planet model for the system. Continued monitoring of the system with more precise measurements may uncover additional planets in the system, including TTV evidence. Follow-up observations could be conducted with the James Webb Space Telescope (JWST) if the system is considered to be a favorable target. To determine whether TOI-1437 b is a strong candidate for atmospheric characterization with JWST, we utilize the transmission spectroscopy metric (TSM) defined by \cite{Kempton2018}. The TSM is proportional to the expected transmission spectroscopy signal-to-noise, based on the strength of spectral features, brightness of the host star, and mass and radius of the planet. The TSM value calculated for TOI-1437 b is determined to be 17. As \citet{Kempton2018} recommends that planets with TSM $>$ 90 for 1.5 $<$ R$_p$ $<$ 10 R$_\earth$ be selected as high-quality atmospheric characterization targets among TESS planetary candidates, TOI-1437 b can be considered as a medium priority target for follow-up observations with JWST. \

Although TESS observations have led to the discovery of planets orbiting solar analogs \citep{Eberhardt2023}, the majority of TESS discoveries have occurred for stars with masses less than solar. TOI-1437 is not a true solar analog, since it is $\sim$10\% more massive and $\sim$17\% larger than the Sun, placing it even further into the tail of the typical TESS exoplanet host distribution. Even so, studying the parameters such as stellar mass, age, and composition of TOI-1437 could enable an improved understanding of how planetary system evolution is affected by these variations in comparison to the characteristics of the Sun \citep{Gaidos1998}. TOI-1437 could thus provide a benchmark for systems that fall close to solar analogs, where the formation, evolution, and architecture of planetary systems around such stars remains difficult to constrain.


\section{Conclusions}
\label{sec:conclusions}

Continued discoveries from TESS are providing increasing insight into the sub-Neptune population of exoplanets, their potential properties, and evolution through time. This work identifies TOI-1437 as a planetary system consisting of a sub-Neptune orbiting a near solar analog, which is a class of host star that is relatively under-represented within the stellar demographic known to host sub-Neptunes. Combining our transit model alongside RV analysis with APF/Levy and Keck/HIRES allowed us to measure the planetary radius and system stellar properties, and to constrain the mass of the planet within 3$\sigma$. TOI-1437 observations with high-resolution near-infrared AO and optical imaging combined with $Gaia$ analysis indicate no evidence for additional stellar companions within the system, allowing us to be confident that our signals are representative of planet TOI-1437 b. Although no other planets have yet been detected within the system, continuation of RV monitoring may reveal planetary companions that improve our understanding of demographics and orbital dynamics in sub-Neptune systems.

\clearpage 
\section*{Acknowledgments}

We thank the time assignment committees of the University of California, the California Institute of Technology, NASA, and the University of Hawaii for supporting the TESS-Keck Survey with observing time at Keck Observatory and on the Automated Planet Finder. We thank NASA for funding associated with our Key Strategic Mission Support project. We gratefully acknowledge the efforts and dedication of the Keck Observatory staff for support of HIRES and remote observing.  We recognize and acknowledge the cultural role and reverence that the summit of Maunakea has within the indigenous Hawaiian community. We are deeply grateful to have the opportunity to conduct observations from this mountain.  We thank Ken and Gloria Levy, who supported the construction of the Levy Spectrometer on the Automated Planet Finder. We thank the University of California and Google for supporting Lick Observatory and the UCO staff for their dedicated work scheduling and operating the telescopes of Lick Observatory. This paper is based on data collected by the TESS mission. Funding for the TESS mission is provided by the NASA Explorer Program. \ 

DP acknowledges support from the NASA FINESST Fellowship issued via Grant no. 80NSSC22K1319. \ 

M.L.H. would like to acknowledge NASA support via the FINESST Planetary Science Division, NASA award number 80NSSC21K1536. \ 

DRC acknowledges partial support from NASA Grant 18-2XRP18\_2-0007. \ 

D.H. acknowledges support from the Alfred P. Sloan Foundation, the National Aeronautics and Space Administration (80NSSC21K0652) and the Australian Research Council (FT200100871). \ 

J.L.-B. was partly funded by the Ram\'on y Cajal program with code RYC2021-031640-I. and by the Spanish MCIN/AEI/10.13039/501100011033 grant PID2019-107061GB-C61. \ 

J.M.A.M. is supported by the National Science Foundation Graduate Research Fellowship Program under Grant No. DGE-1842400. \ 

This research has made use of the Exoplanet Follow-up Observation Program (ExoFOP; DOI: 10.26134/ExoFOP5) website, which is operated by the California Institute of Technology, under contract with the National Aeronautics and Space Administration under the Exoplanet Exploration Program. \ 

This paper includes data collected by the TESS mission that are publicly available from the Mikulski Archive for Space Telescopes (MAST). Data was taken from the TESS Input Catalog \citep{10.17909/fwdt-2x66}. \ 

Some of the observations in this paper made use of the High-Resolution Imaging instrument ‘Alopeke and were obtained under Gemini LLP Proposal Number: GN/S-2021A-LP-105. ‘Alopeke was funded by the NASA Exoplanet Exploration Program and built at the NASA Ames Research Center by Steve B. Howell, Nic Scott, Elliott P. Horch, and Emmett Quigley. Alopeke was mounted on the Gemini North telescope of the international Gemini Observatory, a program of NSF’s OIR Lab, which is managed by the Association of Universities for Research in Astronomy (AURA) under a cooperative agreement with the National Science Foundation on behalf of the Gemini partnership: the National Science Foundation (United States), National Research Council (Canada), Agencia Nacional de Investigación y Desarrollo (Chile), Ministerio de Ciencia, Tecnología e Innovación (Argentina), Ministério da Ciência, Tecnologia, Inovações e Comunicações (Brazil), and Korea Astronomy and Space Science Institute (Republic of Korea).

This research was carried out in part at the Jet Propulsion Laboratory, California Institute of Technology, under a contract with the National Aeronautics and Space Administration (80NM0018D0004). 

The authors thank Eric Mamajek for helpful discussions that improved the quality of this manuscript.

\bibliography{references}
\bibliographystyle{aasjournal}
\end{document}